# LAYPEOPLE'S AND EXPERTS' RISK PERCEPTION OF CLOUD COMPUTING SERVICES

Gianfranco Elena and Christopher W. Johnson

Department of Computing Science, University of Glasgow, Glasgow, UK

## ABSTRACT

*Cloud computing is revolutionising the way software services are procured and used by Government organizations and SMEs. Quantitative risk assessment of Cloud services is complex and undermined by specific security concerns regarding data confidentiality, integrity and availability. This study explores how the gap between the quantitative risk assessment and the perception of the risk can produce a bias in the decision-making process about Cloud computing adoption.*

*The risk perception of experts in Cloud computing (N=37) and laypeople (N=81) about ten Cloud computing services was investigated using the psychometric paradigm. Results suggest that the risk perception of Cloud services can be represented by two components, called "dread risk" and "unknown risk", which may explain up to 46% of the variance. Other factors influencing the risk perception were "perceived benefits", "trust in regulatory authorities" and "technology attitude".*

*This study suggests some implications that could support Government and non-Government organizations in their strategies for Cloud computing adoption.*

## 1. INTRODUCTION

Cloud Computing is considered as one of the key innovation of the 21[st] Century, and its diffusion is expected to generate substantial direct and indirect impacts on economic and employment growth in the EU [1]. For the period 2015-2020, it is estimated a possible cumulative impact up to €250 billion GDP and 3.8 million new jobs. According to International Data Corporation (IDC), worldwide public IT cloud services spending will reach €52.9 billion in 2014 and grow to more than €118.7 billion in 2018 [2].

It is expected that the cloud services market will provide innovative solutions for industry-focused platforms that will revolutionise how companies and organizations operate their IT and how they compete in their own industry [3].

In 2010, the UK Government was the first to introduce a Cloud strategy but despite the adoption of the "cloud first policy" the IT spending on cloud services in 2014 was €24.7 million in comparison to an overall IT Government spending of €5.89 billion.

Cloud computing services have a major potential to bring numerous benefits but some barriers, especially concerning cybersecurity and data protection [4], could limit the economic and innovation advantages in the public sector.

The adoption of Cloud computing does not depend only on technological advances and favourable economic conditions [1] but also on the risk perception and the risk attitude of decision makers like Government officers and IT risk managers [5]. In this exploratory study, we investigate the risk perception of Cloud computing services to understand factors that influence laypeople's and experts' evaluations. Cloud services were described in short scenarios, and participants assessed the perceived risks and benefits associated with these cloud applications. These assessments may identify the Cloud services that are related to higher perceived risks. Furthermore, results should indicate which factors have the greatest impact on the risk perception of Cloud computing services.

Quantitative risk assessment is generally assumed to quantify objectively the likelihood and consequences of adverse events, and it has an important role in the decision-making process to adopt Cloud computing  [6]. We argue that it is also important to understand how people judge and evaluate the hazards that they might be exposed to. The perceptions of risk can influence risk acceptability and behavioural decisions. People's risk perceptions are determined by hazard features and personal risk attitudes. People have different risk attitudes according to how they evaluate a risk situation. Also, "risk attitudes are neither necessary stable nor homogenous across hazard types" [7]. On one hand, people that are risk averse are less likely to take risk and foster digital innovation, on the other hand, people that are willing to take risk could overestimate the risk appetite of their organization.

We argue that risk attitude and risk perception may differ from quantitative risk assessment and can have an impact on the risk acceptability and adoption of Cloud computing services.

Previous studies examined public risk perception of emerging technologies as opposed to specific products [8], [9]. Therefore, we expected that the type of Cloud service would influence the risk perception of the participants. We adopted the psychometric paradigm, as one of the most popular approaches to the study of variations in risk perception across specific hazards. In our study we also investigated determinants of differences in risk perception looking at risk attitude towards technology, perceived risks, perceived benefits, trust in regulatory authorities and gender [10], [11], [12], [13].

This paper has the following structure: first, we review the literature introducing the hypothesis to test. Second, we present our methodology to investigate laypeople's and experts' risk perception. Next, we present our analysis of the results and a discussion of our findings. We conclude with the shortcomings of our work and the future research directions.

## 2. CONCEPTUAL DEVELOPMENT: THE RESEARCH MODEL AND HYPOTHESES

The perception of risk has been a focus of interest of researchers for some decades.  In the literature two main approaches have been proposed to explain risk perceptions: the Psychometric Paradigm, as developed by psychologists [14],[15] and Cultural Theory proposed by anthropologists and sociologists [16].

The Psychometric Paradigm is based on the assumption that some characteristics of risks are perceived similarly (e.g. voluntariness is correlated with controllability, observability with knowledge about the risk). Based on the correlation between some of these risk characteristics – usually called "items" – they can be combined into two or three components using multivariate and factor analyses. Each component thus consists of several highly correlated items. Former risk perception studies [17] typically identified two to three components named: "dread risk", "unknown risk" and "trust" [15].

"How safe is safe enough?", published by Fischhoff and Slovic in 1978, was one of the first psychometric studies of attitudes towards technological risks and benefits.  The study used eight characteristics to measure three underlying dimensions of  people's risk perception of different hazards [14].  Psychological researchers have identified 47 known items that influence people's perception of risk (e.g. Voluntariness of exposure, Controllability of consequences, Knowledge of the risk etc.) [18],[19]. The study of risk perception was considered in many areas, such as nuclear engineering [20],[21], epidemics [22], automobile safety [23] and construction safety [24]. In the field of information technology, some efforts have also been made. Other authors [25] reviewed the social science literature on the public perception of risk and extended their discussion to perceptions of crime in cyberspace. Vyskoc and Fibikova  conducted a survey about how IT users' perceive information security [26].

The psychometric approach has been used to describe how laypeople and experts judge risks [17]. This is because "laypeople" and "experts" often define risks differently. These differing conceptions result in laypeople assigning relatively little weight to risk assessments conducted by technical experts [27-30]. "Experts", making frequent use of statistical data to evaluate risks and benefits, are assumed to be able to provide an objective risk assessment. Nevertheless, an expert is a specialist in a specific area. It is unlikely he can be experienced on all the topics. Hence, if the experts have to evaluate a new technology they can be forced to rely on intuition and extrapolation resulting in a risk assessment similar to those of the laypeople [17],[31-33].

Laypeople's perception of risk is highly correlated to the component called "dread risk" which measures general risk of a new technology capturing the severity, the probability and the riskiness. The higher the risk topic is judged on this factor and the more people want to see its current risks reduced and regulated [15]. In those cases in which the risk topic ranks high on the "dread risk" and "unknown risk" dimensions that issue is very likely to be discussed in the general public and in the media. On the other hand, risks that rate low on the dimensions of "dread" and "unknown" risk are often underestimated by the general public. Lifestyle risks were the largest category reported to be neglected [29]. "Studies using the psychometric paradigm have shown that it is possible to quantify and predict perceived risk. The technique seems appropriate to identify similarities and differences in the perception of different risks" [17]. Following this argumentation, we argue that investigating the risk perception of Cloud computing can provide useful information to forecast public acceptance. For example, if a previously unsearched topic, like "storing data in the Cloud", is judged similarly as a previously researched topic, like "using mobile phones", then we can expect similar constraints in the discussion of stakeholders.

Laypeople use speculative frameworks to make sense of the world and selective judgement in their  responses to risk [30]. They usually balance their lack of knowledge employing social trust when assessing the risks of a new technology [10]. Research in the various domains showed that people who trusted regulating authorities attributed more benefits and fewer risks to this technology [31],[12]. General attitudes toward technology are likely, therefore, to influence the assessment of Cloud computing technology. People who are positive toward technology will likely assess Cloud technology applications more positively than other people. Similarly, negative correlations between perceived benefits and risks have been observed [10]. Then, we assumed that laypeople would assess perceived risks taking into consideration perceived benefits. People who perceive more benefits are likely to perceive less risk.

"Experts and laypeople perceive risks differently but it is not clear whether the observed differences are due to expertise or to other factors" [32],[33],[34] [35]. We assume that experts' risk assessments are prevalently based on their knowledge about a technology and not on perceived benefits or attitudes.

Several studies have demonstrated that the psychometric approach has a stronger explanatory power [36],[37] even if sociologists reject that perception of risks and public acceptability of technologies can be reduced to a limited number of qualitative risk characteristics [38]. They also argued that individual variations are in large part due to the cultural groups to which we belong [39], and that "risk perception investigations should be accompanied by social, cultural, and political analyses in order to understand their deeper implications" [40] [41]. Numerous empirical investigations have been carried out on the cultural theory [42],[43],[44],[45] and the psychometric paradigm [46],[47],[48],[49] and a number of comparative analyses have been performed [37],[36],[40] to investigate the power with which the psychometric risk characteristics can actually explain risk perceptions.

In particular, "all these studies contributed to explain the explanatory power that each method can claim to account for, revealing their limitations" [50]. As reported by Sjöberg [36], it is commonly accepted among the risk community that the psychometric approach is able to explain a greater proportion of the variance in perceived risks, than the cultural theory [51].
In this context, we decided to investigate laypeople and experts' risk perception of Cloud computing applications using the psychometric paradigm and we tried to account for other factors influencing participants' risk ratings.

First, we present the overall perceived risk of ten cloud applications and three non-cloud applications for laypeople and experts. Second, we define the risk components that can predict the risk perception drawing two maps of the risk perception for Cloud computing services. Finally, we investigate some of the factors influencing the risk perception.

According to the previous considerations we formulate the following hypothesis:

- H1. The perceived risk of Cloud services is different
- H2. "Dread Risk" affects "Risk Perception"
- H3. "Knowledge of Risk" affects "Risk Perception"
- H4. "Perceived Benefit" affects "Risk Perception"
- H5. "Technology Attitude" affects "Risk Perception"
- H6. "Perceived Trust in regulatory authorities" affects "Risk Perception"
- H7. "Gender" affects "Risk Perception"

# 3. RESEARCH METHODOLOGY AND DATA COLLECTION

## 3.1 Participants

An on line survey was distributed to 285 potential participants in the UK and a total of 118 people took part in the research. Acceptance rate was 41%. A pilot test of the questionnaire was administered to 15 university students to identify errors, avoid wrong design and predict possible problems. 78% (N=92) of the respondents were male and 22% (N=26) were female. The mean age was 38 (SD = 10.9). Participants were chosen between IT Cloud experts and laypeople. An MCAR test was used to test missing values at random. The expected maximization function was used to complete missing values (<5%) [52].

IT Cloud experts' sample fulfils the following criteria: more than two years working experience in the field of Cloud computing, using 5 or more Cloud services at least twice a week, knowledge of Cloud services (=>4 on a 5-points Likert scale), more than three years IT experience. A total of 37 experts with a mean age of 37 (SD = 11.26; ranging from 24 to 65) took part in the survey. We claimed that "expert" participants had a deeper knowledge of Cloud computing services based on the selection previous mentioned selection criteria, but we do not make any case about the accuracy of their knowledge.

The laypeople's sample size was 81 and the mean age was 38 (SD = 10.84; ranging from 18 to 62).

## 3.2 Questionnaire

The survey aims to get a measure of the risk perception of ten Cloud computing services. Each Cloud service was described in a short scenario in order to facilitate its comphreension. understand and specific to the Cloud computing domain.

Since many different technical architectures are referred to as "the Cloud", we provided a brief explanation of Cloud computing to standardize the definition and the meaning of Cloud computing technology: "Cloud computing is a model for enabling ubiquitous, convenient, on-demand network access to a shared pool of configurable computing resources (e.g., networks, servers, storage, applications, and services) that can be rapidly provisioned and released with minimal management effort or service provider interaction" [53].

Also, information about generalized risks of using Cloud services was summarized in an introductory section of the questionnaire: "Cloud computing services are associated with new risks. Sensitive data are processed outside the enterprise and that brings to bypass the physical, logical and personnel controls that Information Technology departments exert over in-house programs. Customers are ultimately responsible for the security and integrity of their own data and they won't know exactly where their data are hosted and how the data environment is shared with other customers. Moreover, any vendor offering that does not replicate the data and application infrastructure across multiple sites is vulnerable to a total failure. Finally, investigating inappropriate or illegal activity may be impossible in Cloud computing and long-term viability be at risk since Cloud provider could go broke or get acquired by a larger company" [50] .

After this introduction, the benefits of 10 Cloud computing services and 3 non-Cloud computing services were briefly described in short scenarios (Appendix 1). We included 3 non-Cloud computing services (mobile phones, wireless devices and usb flash drives) to obtain baseline values for Cloud perceived risks. We considered "mobile phones" because they are commonly used to exchange and store personal data, "wireless devices" as a means to connect to internet resources wireless, "usb flash drives" as one of the most used way to exchange data files and applications.

Participants were asked to rate the hazards on a 5-point Likert scale. The variables utilized in earlier psychometric studies that also took a scenario-based approach [14],[15],[55],[56],[34] [56] were adapted for the examination of Cloud computing hazards. Some of the scales (e.g., old-new hazard) could not be used and were replaced by other scales (e.g., trust). The terms used for the characteristics were not explained explicitly to the participants because the relevant constructs are the eight risk characteristics and subjective interpretation of said characteristics is appropriate [38]. Questions and rating scales used are reported in appendix 2.

# 4. RESULTS

Results show some differences in the risk perception of laypeople and experts, and contribute to identify other factors influencing the risk perception of Cloud computing services.

## 4.1 Perceived Risks

The mean levels of perceived risk were compared between 10 cloud services and 3 non-cloud services. The mean values and standard deviation for perceived risks of laypeople and experts are given in Table 1 and Fig.1. An independent samples one-way analysis of variance showed that there is a significant difference between the level of perceived risk across the cloud and non-cloud services for laypeople $F_{(12,1039)}=2.9$, $p<0.001$ and IT Cloud experts $F_{(12,468)}=2.03$, $p<0.05$. This confirms hypothesis H1. The items "Storage and Backup", "Documents Sharing" and "Identity and Access Management" received the highest risk rating in both samples. Results show that on average participants considered using Cloud services as risky as using external hard drives and mobile phones. Some cloud services (e.g. "Digital Marketplace", "Computing Power") were considered less risky than non-cloud services. Results show that laypeople perceived mean values of perceived risk higher than IT Cloud experts across the majority of the items presented.

Table 1. Descriptive statistic of survey responses (N=118). * Non-Cloud Applications.

| CLOUD COMPUTING APPLICATIONS | LAYPEOPLE N=81 | | EXPERTS N=37 | |
|---|---|---|---|---|
| ITEMS | MEAN | SD | MEAN | SD |
| 1. Storage and Backup | 3.050 | 0.967 | 2.892 | 1,022 |
| 2. Documents Sharing | 2.963 | 0.858 | 2.946 | 0,780 |
| 3. Identity and Access Management | 2.926 | 0.905 | 3.054 | 0,941 |
| 4. Software Applications as Service | 2.840 | 0.901 | 2.595 | 0,956 |
| 5. Email system management | 2.815 | 0.896 | 2.541 | 0,730 |
| 6. Digital Marketplace | 2.519 | 0.950 | 2.514 | 0,804 |
| 7. Virtual Private Cloud | 2.617 | 0.982 | 2.568 | 1,015 |
| 8. Monitoring and Auditing | 2.580 | 0.804 | 2.432 | 0,867 |
| 9. Computing Power | 2.605 | 1.080 | 2.459 | 1,043 |
| 10. Vulnerability Assessment | 2.679 | 0.739 | 2.351 | 0,633 |
| *11. USB pens and External HD | 2.951 | 0.973 | 2.649 | 1.086 |
| *12. Mobile Phones | 2.926 | 0.919 | 2.865 | 1,004 |
| *13. Wiress Devices | 2.815 | 0.896 | 2.703 | 1,102 |

## 4.2 Analysis of the Aggregated Data of the Laypeople Sample

To explore the latent structure of the 8 items and identify the factors that can influence laypeople's perception of Cloud computing, an exploratory factor analysis was conducted on the raw data. The sample was reliable, as Cronbach's Alpha was 0.802. The 8 items resulted suitable to run a factor analysis (value=0.653) based on the Kaiser-Meyer-Olkin (KMO) measure of sampling adequacy. Also, Bartlett's Test of Sphericity (value  50.42, p<0.01) resulted acceptable [57].

The aggregated data of the laypeople sample were submitted to a principal component analysis with varimax rotation. Since data outliers influenced correlation coefficients, we computed rank correlations among the eight rating scales and a principal component analysis of these rank correlations. Based on the scree-test plot, two components, accounting for 66% of the variance, explained the correlations among the eight rating scales [50]. The first component, labelled as "dread risk" in accordance with previous studies [15], measures general risk capturing the severity, the probability and the riskiness of a Cloud computing service. The second dimension, called "unknown risk", indicates the novelty of the risk measuring both knowledge, justifiability and voluntariness characteristics of the perceived risk. As Table 2 shows, the first of the two orthogonal components of the rotated factor loadings is highly correlated with perceived "probability of IT security incidents", "adverse security effects" and "worries about the risks". The second component is positively associated with the variables "unknown risk" and "justifiability of taking the risk" while it is negatively associated with the variable "voluntariness". This component is labelled "unknown the risk" [58],[14],[15].

Table 2, 3. Results of the Principal Components Analysis show the loadings for the eight rating scales, averaged across individuals, for the Laypeople's and Experts' Sample

| Laypeople Rotated Component Matrix | | | Experts Rotated Component Matrix | | |
|---|---|---|---|---|---|
| Rating scale | Components | | Rating scale | Components | |
| | Dread Risk | Unknown Risk | | Dread Risk | Unknown Risk |
| Prob_Incident | **.958** | .014 | Prob_Incident | **.818** | .434 |
| Adv_Effects | **.931** | -.065 | Adv_Effects | **.808** | .403 |
| Risk_Worries | **.930** | .038 | Risk_Worries | **.688** | .389 |
| Justifiable | .481 | .474 | Justifiable | .506 | -.209 |
| Knowledge | .164 | **.857** | Knowledge | -.099 | **.878** |
| Voluntariness | .096 | **-.767** | Voluntariness | .558 | **-.642** |
| Trust | -.432 | .597 | Trust | -.104 | -.436 |
| Control | .321 | .356 | Control | .165 | **.831** |

Rotation Method: Varimax with Kaiser Normalization.
Loadings exceeding 0.6 are in boldface.

To test how the factor scores, related to the two principal components, influenced perceived risks we performed a linear regression analysis. The statistical analysis software (SPSS) was used to enter simultaneously all the indipendent variables and compute the regression analysis. Inspecting possible outliers suggested that a transformation of the response variable was not necessary. The proposed model was significant (F (2,10)=13.32, p<0.01) and explained 46% of the variance in perceived risks. The first component, "dread risk," was the most important predictor ($\beta$ = 85, p < 0.001). The second component "Unknown Risk" was less significant ($\beta$ = 26 p < 0.01). The third component "Trust in regulatory authorities" was not significant. These results confirmed hypothesis H2 and H3. The use of the two components, "dread risk" and "unknown risk" as coordinates [59], allowed us to represent the risk perception of ten Cloud computing services and three non-Cloud technologies on a two-dimensional plot. Results show that the risk perception of Cloud computing services varies according to the type of service. Cloud services with higher loading on the component "dread risk" and "unknown risk" were: "storing and backup data in the Cloud", "creating and sharing documents in the Cloud", "identity and access management", "software applications as a service" and "monitoring and auditing". These cloud services may be the most likely to be targets of public discussions about the adoption Cloud computing services. Accidents with these characteristics are seen as "high-signal accidents" causing high societal impact, far higher than the direct immediate costs of the accidents [15]. Participants perceived the first three cloud services as riskier than using their mobile phones to exchange messages and photos. The use of USB flash drives is perceived very risky but it is associated with a low level of unknown risk. The risk perception of all the other cloud services was quite low. A higher raking on the "unknown risk" component may suggest a poor knowledge of the risks involved. The cloud service "digital marketplace" ranked as the less risky.

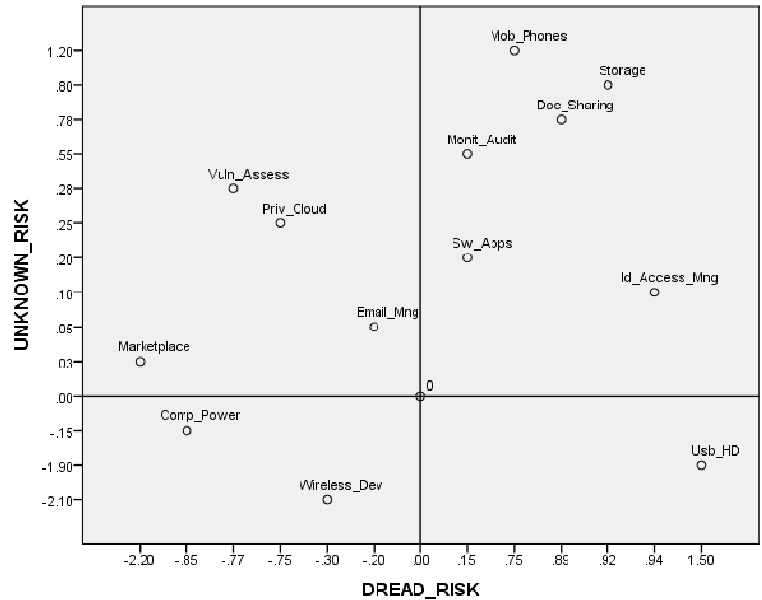

Fig. 2. It presents the risk perception of laypeople for 10 Cloud computing services and 3 non-cloud technologies, using the factor scores of "Dread Risk" and "Unknown Risk" as coordinates.

## 4.3 Analysis of the Aggregated Data of the Expert's Sample

The sample resulted reliable with a Cronbach's Alpha value of 0.81. Kaiser-Meyer-Olkin Measure of Sampling Adequacy was 0.678 and Bartlett's Test of Sphericity was 51.88 p<0.001. The aggregated data of the IT Cloud experts sample were submitted to a principal component analysis with varimax rotation. Since data outliers influenced correlation coefficients, we computed rank correlations among the eight rating scales and a principal component analysis of these rank correlations. Based on the scree-test plot, two components, accounting for 71.96% of the variance, explained the correlations among the eight rating scales [50]. As represented in table 3, the first orthogonal components labelled "dread risk" is highly correlated with "perceived probability of IT security incidents", "adverse security effects" and "worries about the risks". The second component is labelled "unknown risk" and it is positively associated with "unknown risk" and "control of the risk".

To test how the factor scores, related to the two principal components, influenced perceived risks, we performed a linear regression analysis. All the indipendent variables were entered simultaneously to compute the regression analysis. Inspecting possible outliers suggested that a transformation of the response variable was not necessary.

The resulting model (F (2,10)=17.64, p<0.01) explained 37% of the variance in perceived risks and it was statistically significant. The first component, called "dread risk," was the most important predictor ($\beta$ = 86, p < 0.001). The second component "unknown risk" was marginally significant ($\beta$ = 21, p <0.01). The third component "Trust in regulatory authorities" was not significant. These results confirmed hypothesis H2 and H3 for the experts sample and allowed to represent the risk perception of Cloud computing services on a two-dimensional plot using factor scores of the two components - "Dread Risk" and "Unknown Risk" - as coordinates. Experts showed an overall lower level of risk perception than laypeople.

Results show that the risk perception of Cloud computing services varies according to the type of service. The cloud service with higher loading on the component "dread risk" and "unknown risk" was "creating and sharing documents in the cloud with other people". Experts perceived this cloud service as riskier than using their mobile phone to exchange messages and photos. The use of USB flash drives is perceived as low risk. The level of risk perception for all the other cloud services was quite low but with high loading on the "unknown risk" component. The cloud service "digital marketplace" ranked as one of the less risky but the high loading on the "unknown risk" component.

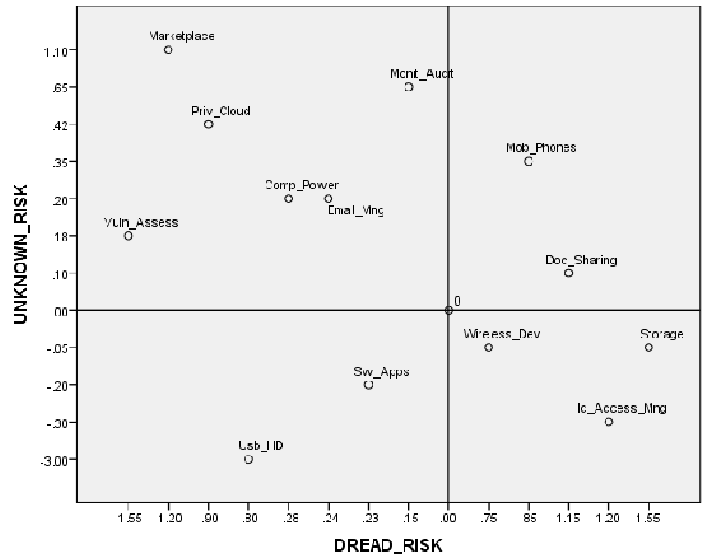

Fig. 3. It presents the risk perception of cloud experts for 10 Cloud computing services and 3 non-cloud technologies, using the factor scores of "Dread Risk" and "Unknown Risk" as coordinates.

## 4.4 Factors Influencing Individual Differences in Laypeople's Risk Perception

According to previous studies, we also hypothesized that "trust in regulatory authorities" and "perceived benefits" influence perceived risks [11]. Moreover, we expected that "technology attitude" and gender would influence the level of perceived risks associated with Cloud computing services [60],[41].

In tables 4 and 5, we represented the correlations between predictors and perceived Cloud technology risks for Laypeople and Experts. Laypeople's assessments of trust, benefit and risks of the different Cloud computing services were highly correlated. We entered all the factors to compute a linear regression analysis. Results, reported in table 6, were partially in line with our hypotheses. "Perceived benefits" and "trust in authorities" influenced significantly perceived cloud risks. Laypeople perceived lower levels of risk associated higher level of benefits from using Cloud applications. The reverse result was observed for technology fears. Perceived Cloud benefits, technology benefits, technology fears and trust in government authorities influenced the risk perception of laypeople. Age and gender were not significant. The regression model explained 38% of the variance in perceived Cloud services risk for the sample of laypeople. These results confirm the importance of building trust in the regulatory authorities and explaining the benefits of using cloud computing applications [6],[4].

Table 4. Correlations between factors influencing the risk perception for Laypeople

|  | 1. PCR | 2. PCB | 3. Trust | 4. TeB | 5.TeF | 6. Gender |
|---|---|---|---|---|---|---|
| 1.Perceived Cloud Risks |  | -0.56* | -0.30* | -0.87** | 0.69** | 0.24** |
| 2.Perceived Cloud Benefits | -0.56* |  | 0.28* | 0.21* | -0.21* | -0.12* |
| 3.Trust in Authorities | -0.30* | 0.28* |  | 0.23* | -0.29* | -0.19* |
| 4.Technology Benefits | -0.87** | 0.21* | 0.23* |  | -0.36* | -0.13* |
| 5. Technology Fears | 0.69** | -0.21* | -0.29* | -0.36* |  | 0.20* |
| 6. Gender | 0.24** | -0.12* | -0.19* | -0.13* | 0.20* |  |

*Note*: * <0.05, ** <0.01

Table 5. Correlations between factors influencing the risk perception for IT Cloud Experts

|  | 1. PCR | 2. PCB | 3. Trust | 4. TeB | 5. TeF |
|---|---|---|---|---|---|
| 1.Perceived Cloud Risks |  | -0.26* | -0.50* | -0.57 | 0.39 |
| 2.Perceived Cloud Benefits | -0.26* |  | 0.32* | 0.37* | -0.11* |
| 3.Trust in Authorities | -0.50* | 0.32* |  | 0.43* | -0.32 |
| 4.Technology Benefits | -0.57 | 0.37* | 0.43* |  | -0.39 |
| 5. Technology Fears | 0.39 | -0.11* | -0.32 | -0.39 |  |

*Note*: * <0.05, ** <0.01

Table 6. Multiple regression analysis with perceived cloud applications Risks as dependent variable for the Laypeople and Experts' samples

| Predictors | Laypeople | | | Experts | | |
|---|---|---|---|---|---|---|
|  | Beta | *t*-value | Hypothesis | Beta | t-value | Hypothesis |
| Perceived Cloud Benefits | -0.25 | -5.34* | H4 confirmed | 0.17 | 1.35 | H4 not confirmed |
| Technology Benefits | -0.13 | -2.78* | H5 confirmed | 0.02 | -0.17 | H5 not confirmed |
| Technology Fears | 0.25 | 5.63* | H5 confirmed | 0.02 | 0.16 | H5 not confirmed |
| Trust in Authorities | -0.37 | -8.24* | H6 confirmed | -0.82 | -5.83 | H6 not confirmed |
| Gender | 0.06 | 1.21 | H7 not confirmed | - | - | N/A |

*Note*: * <0.05, ** <0.01; $R^2$ (Laypeople) = 0.38*; $R^2$ (Experts) = 0.32*

## 4.5 Participant Feedback

In order to support our findings, participants were asked to describe the risks related to the use of a cloud service to create and share unclassified working documents. Results, as reported in table 7, suggest the most important perceived risks associated to the use of Cloud services in their work environment. These results are in line with previous studies [6].

Table 7. Main perceived risks associated to the cloud scenario "Document sharing". (N=118)

| MAIN RISKS | % of sample (N=118) | RISK LEVEL |
|---|---|---|
| Loss of Confidentiality | 27.1 % | High |
| Loss of Control of services and/or data | 20.3 % | High |
| Loss of Privacy | 17.8 % | High |
| Loss of Availability of services and/or data | 8.5 % | High |
| Loss of Integrity of services and/or data | 7.6 % | High |
| Lack of Reliability | 5 % | High |
| Lack of Liability of providers in case of security incidents | 4.2 % | High |
| Loss of Reputation and Credibility | 3.4 % | High |
| Uncontrolled variable cost | 2.5 % | High |
| Intra-Clouds (vendor lock-in) migration | 1.7 % | High |
| Cost and difficulty of migration to the Cloud | 0.85 % | Medium |
| Inconsistency between trans national laws and regulations | 0.85 % | Medium |

# 5. DISCUSSION

Cloud computing represents a paradigm shift in the way software and infrastructure services are delivered. As part of an European digital innovation strategy, the adoption of Cloud computing services could support an improvement of business efficiency and flexibility [1] [61]. In this study we suggested that factors influencing the risk perception of Cloud computing should be considered to support Cloud computing adoption's strategies. How people perceive the risks and benefits associated with cloud computing services is very important to enhance effective adoption strategies.

In this study we investigated four main questions: how laypeople and experts perceive the risk of cloud computing services? What is the difference in the risk perception of each cloud service? Which factors contribute to influence the risk perception of laypeople and IT cloud experts? What are the perceived risks for creating and sharing an unclassified document in the Cloud?

In the first part, results suggested that the risk perception of cloud computing services varies according to the type of service and that laypeople have a higher level of risk perception than experts, as they probably use different cues to assess the risks of cloud services. It is not clear what are the factors influencing these differences [32],[33] but it seems that experts based their risk assessments more on the knowledge of the risks than on the perceived benefits or personal attitude [35]. Another difference could be that laypeople generally focus on the magnitude of the risk rather than on its probability. It is likely that, when laypeople evaluated cloud risks, the possible consequences of that scenario were considered more important than the low probability that it could happen [62].

In the second part, we adopted the psychometric paradigm to investigate the risk perception of ten cloud computing services and three non-Cloud technologies, which we used as a baseline to identify similarities and differences in the perception of different risks [15]. Results of the principal component analysis and regression analysis indicated that two components, called "Unknown Risk" and "Dread Risk", explained most of the variance of perceived risks. The relations between these components and the perceived overall risk of Cloud services were found and then tested by multiple regressions. The results confirmed that two factors could explain up to 46% of the variance for the laypeople and up to 37% of the variance for the experts' sample.

On the one hand, as represented in figure 2, laypeople perceived the three most dreaded cloud services as riskier than using their mobile phones to exchange messages and photos. People that use their mobile phones frequently perceive lower risks than people who use their mobile phones infrequently [63]. Then, as millions of people use their mobile every day, we assumed that mobile phones may give an indication of an average level of the acceptable risk. The cloud services with higher loading on the components "unknown risk" and "dread risk" are frequently overestimated risks [51] and the most likely to be enhanced by a social amplification process [64]. It appears that laypeople have limited knowledge of the risks and benefits of using cloud services. Providing a better understanding of the opportunities and security measures of the cloud services could support their adoption.

Another result is that laypeople perceived the use of USB flash drives as very risky but associated with a low level of "unknown risk". A recent survey data indicated that the portability of USB flash drives represents a significant risk of data loss [65]. Another research found that about twelve percent of corporate end users reported finding a flash drive in a public place which retained a significant amount of identifiable information [66]. A last point is about the cloud service "digital marketplace" that ranked as the less risky. We assumed that the rigorous procedures applied by the most well-known cloud stores to select and approve their software products [67],[68] played an important role in this evaluation. It seems plausible to suggest that the implementation of trustworthy cloud infrastructures is a keystone to foster cloud computing services adoption.

On the other hand, as represented in figure 3, IT cloud experts perceived "creating and sharing documents in the cloud with other people" as the cloud service with higher loading on the components "dread risk" and "unknown risk". The high level of risk perception about using "a mobile phone to exchange data and messages" suggests that there is a limited knowledge of the related risks. Another difference with laypeople is that the use of USB flash drives is perceived as low risk. This result suggests that experts have a very good knowledge of the risks and know how to manage them. The overall level of risk perception for the other cloud services was quite low. The cloud service "digital marketplace" ranked as one of the less risky but with a high loading on the "unknown risk" component. This may suggest that experts have "raising concerns about how third-party applications may misuse or improperly handle users' privacy-sensitive data" [69]. Cloud service providers use very sophisticated techniques to control and publish third party applications on their cloud stores, but "their tests improves the chances of detecting potential malware, but none of them can prove that no malware exists" [70].

In the third part, we accounted for other factors influencing the risk perception of cloud services. Factors influencing the risk perception of laypeople were: "perceived benefits", "trust in the regulatory authorities" and "technology attitude". The related regression model could explain an additional 38% of the variance in the risk perception.

Laypeople have a little knowledge of cloud risks and they may be influenced by their belief and values in the evaluation of new technologies [51]. We argued that perceived benefits may reduce the risk perception of cloud services. Therefore, laypeople who perceive more benefits are likely to perceive less risk. Public concerns and risk issues about using cloud computing are reduced if people are familiar with the benefits of cloud computing. It remains not clear the role of heuristics and probability judgement biases as well as the influence of mass media, which is still under debate [60],[71].

The importance of trust for risk perception has been demonstrated in numerous studies [31]. The present research confirms that "trust in the regulatory authorities" is an important predictor of the risk perception of cloud computing services. We argue that "trust in the regulating authorities" may support laypeople to balance their lack of knowledge as well as to perceive higher level of benefits and fewer risks to cloud services [10], [12], [31].

"Technology attitude" was another important factor influencing the risk perception of cloud services. "Technology benefits" reduced the perceived risk and improved the perceived benefits while "technology fears" augmented the perceived risk and decreased the perceived benefits. We argue that laypeople's attitudes toward technology are likely to influence the assessment of cloud computing services. Laypeople may also be influenced by their belief and values in the evaluation of cloud services [51].

In the fourth part, participants indicated the main risks about "creating and sharing an unclassified document" using the Cloud (table 7).

The "loss of privacy" and "loss of confidentiality" could potentially have a high impact on the reputation of the organization. On the one hand, system administrators and auditors, working in the cloud providers, could become targets for criminal organizations. On the other hand, in public clouds, errors or attacks could enable an attacker to access the resources of one or more specific customers of the cloud service provider.

The "loss of control of services and data" could have a potentially severe impact on the organization's capacity to meet its mission and comply with the security policies. The use of a cloud infrastructure implies to not being able to implement and manage some security measures like vulnerability assessments and penetration testing. Also, cloud providers may outsource some services to unknown providers which may be not compliant with the requested security standards. The "loss of availability" could be determined by a loss of Internet connectivity, a technical problem of the cloud service provider or a reduction of the network bandwidth. It could have a high financial and reputational impact on the organization [4].

In summary, it is hard to account for the risk perception of cloud computing services because it presents multidimensional factors and their influence it is not clear. We argued that laypeople and experts have a different risk perception of cloud services and that the psychometric paradigm can be modified to investigate these differences [32],[33]. We investigated other factors which have a significant influence on the risk perception but many other factors remain to be searched and explained.

## Implications for the European and National Government Agencies

- should promote actions (e.g. conferences, industry trade shows, media campaign) to enhance the public knowledge of Cloud computing. People should be aware of the advantages and limitations of Cloud computing services;

- should improve the "trust in the regulating authorities" by: preventing any incident in the Cloud with significant negative consequences and unwanted side effects; promoting a transparent mechanism to report incidents in the Cloud service providers [72]; defining clear policies at the European and national level;
- should take in consideration that laypeople may be less willing to adopt Cloud computing services. Government innovation strategies should be based not only on quantitative risk assessments [6] but also on the public attitude and risk perception;
- should consider the role of the risk perception in the development of Cloud computing adoption strategies;
- should align the risk and reward for adopting Cloud services so that incentive structures will improve the decision making and risk management process. In fact, the distribution of benefit across the procuring administration does not match the risk distribution, those benefiting most might not be those bearing the highest risk;
- should further investigate how risk perception affects the risk acceptance of Cloud computing services. This could help to develop better adoption strategies and allow Cloud users to take secure behaviours.

**Implications for the Cloud Industry**

- should communicate effectively the benefits and opportunities offered by new cloud services;
- should ensure with timely information that they constantly manage and solve security issues;
- should allow to visualise information about security threats;
- should provide more information on how they protect data in the Cloud from cyber-attacks or infrastructure's incidents;
- should support the European Cloud computing incident reporting system to reduce the risk of any significant negative event and unwanted side effects;
- should consider the risk perception and the public attitude as an important factor to support marketing strategies [6],[4];
- should promote further research on the risk perception and risk acceptance of Cloud computing services.

# 6. LIMITATIONS AND FUTURE RESEARCH

This research presents some limitations. It provides a snapshot of how laypeople and IT Cloud experts perceive the risk of using some Cloud computing services in their work environment. However, risk attitudes toward Cloud computing are not static. The actions of Cloud service providers and the manner in which the media report on Cloud computing can influence risk perception of this technology.

Because the term "Cloud" can be misrepresented, we gave a standard definition of Cloud computing and described the applications in some detail. We addressed possible risks associated with Cloud computing services in the introductory section of the questionnaire because the risks associated with the various applications are similar. Benefits associated with the applications, on the other hand, were mentioned in the short scenarios describing each of the Cloud computing services. Then, it is possible that "people assessed the applications more negatively when benefits were less salient and possible risks of Cloud computing applications were more salient" [50].

However, we would expect that this could influence the mean values, not the observed associations or patterns .

Another potential drawback within this study regards the risk rating scales adopted. Although the scales used have been extensively validated and applied as part of the psychometric paradigm, over a 25-year history [38], more components could have been used and some of the scales do not necessarily fit well with the questions asked and directly associated with them.

Moreover, despite the results that we have presented, the size of our sample likely affected the possibility to infer more about the role of the factors contributing to risk perception in the experts' sample. Finally, since we used the psychometric paradigm as a research framework, some limitations of this approach should also be considered [59].

First, principal component analysis of a matrix with only 8 items is bound to give few factors. Second, the scales were based on eight variables suggested in the earlier risk literature but, although 8 scales may seem like a lot, they are likely missing important factors. Third, the finding that a very large share of the variance of perceived risk could be explained by these factors is due to the fact that we analysed mean ratings and not raw data. Mean data are less subject to error than raw data [51].

In the end, future studies may: use other research paradigms to investigate public risk perception of Cloud computing services; investigate the difference in public risk perception between SMEs, the industry and governmental agencies; explore the relationship between risk perception and risk acceptance of Cloud computing technology; and try to understand how the risk perception and the risk attitude of cloud computing change after an incident in the cloud.

## APPENDIX 1 – SCENARIOS FOR CLOUD COMPUTING SERVICES

The following list of stimuli and descriptions were used for the samples of laypeople and experts.

1. **Storage and Backup**. When you connect to your Cloud provider the Storage Service provides a fully redundant data storage infrastructure for storing and retrieving any amount of data, at any time, from anywhere on the Web. All sorts of important data (e.g. letters, excel files, photos, contacts, databases etc...) on your PCs and mobile devices can be automatically backup daily.

2. **Documents sharing**. Using Cloud services you can create and share with other people any office unclassified document on the Cloud from all your mobile and fixed devices.

3. **Identity and Access Management**. You can securely control access to your Cloud applications (i.e. email, office suite, data storage). You can create and manage security credentials (i.e., access keys, password, Multi-Factor Authentication devices).

4. **Software Applications as a Service**. Software applications can be downloaded from the Digital Marketplace to any of your PCs, Tablet or Smartphone. You don't need to install or update new versions of software and you don't have to worry about keeping your devices in sync. If you delete an email, add a calendar event, update a contact or add a bookmark, the Cloud makes all your changes everywhere.

5. **Email System Management.** Cloud services allow you to manage your emails more efficiently, connecting from any device, blocking more than 99% of spam, viruses, attacks and risky sites, protecting you before the threats reach your network.

6. **Digital Marketplace** is an online store where you can find, download, and immediately start using software that runs on the Cloud (e.g. Apple Market, Google Apps Market etc...). It is an online catalogue of Cloud services and contains details of each of the suppliers and their services. You can use the Marketplace to select the services that best suit your needs.

7. **Virtual Private Cloud** is a service which allows customers to have dedicated Cloud servers, virtual networks, Cloud storage and private ID addresses. The Virtual Private Cloud infrastructure is managed by a public Cloud vendor but the resources allocated to a VPC are not shared with any other customer.

8. **Monitoring and Auditing.** Auditing and Monitoring Service allows detecting suspicious behaviours by external users or employees, or malfunctions on Cloud resources. You can access up-to-the-minute statistics and view graphs. You can monitor for specific events or performances. You no longer need to set up, manage, or scale your own auditing systems and infrastructure.

9. **Computing Power**. In the Cloud you can easily and cost-effectively process vast amounts of data. For instance, if you need to run an experiment and you know that it will take 1000 working hours on your PC to complete all the tasks, then you can decide to rent 1000 Virtual PCs in the Cloud and conclude your experiment in just one hour. Elastic Compute Cloud delivers scalable and pay-per-use compute capacity in the Cloud.

10. **Vulnerability Assessment** service helps you to secure and protect your Cloud resources through accurate vulnerability scanning and actionable reporting. It provides transparent objectivity through continual comparison of your status against widely adopted data security standard.

11. **USB pens and external hard drives** allow you to store and backup data (e.g. letters, excel files, videos, photos, contacts, databases etc...) *(Non Cloud Application)*

12. **Mobile phones** allow wireless voice and data communication. You can exchange short messages, emails and photos with all your contacts *(Non Cloud Application)*

13. **Wireless devices** can be easily relocated with no additional cost of rewiring and without any disruption to the service (eg. Laptops, Printers, Keyboards, Remote Controller, Access Routers). *(Non Cloud Application)*

### APPENDIX 2 - QUESTIONS AND RATING SCALES [50]

| | |
|---|---|
| 1. What is the probability of IT security incident for your organization? | (1 = very improbable; 5 = very probable) |
| 2. Are you worried about risks for your organization? | (1=not worried; 5=very much worried) |
| 3. Do people take the risk voluntary? | (1 = voluntary; 5 = involuntary) |
| 4. Do people know the risk they are exposed? | (1 = known precisely; 5 = not known) |
| 5. How do you rate adverse security consequences for your organization? | (1 = not at all; 5 = very strong) |
| 6. How do you assess control over risk? | (1 = controllable; 5 = uncontrollable) |
| 7. How much do you trust in governmental agencies responsible for protecting people IT security? | (1 = no trust; 5 = much trust) |
| 8. Using this technology is ethically justifiable to foster innovation ? | (1 = not justifiable; 5 = absolutely justifiable). |
| 9. How beneficial do you consider this item to be for your organization as a whole? | (1 = very low; 5 = very high). |
| 10. How risky do you consider this item to be for your organization as a whole? | (1 = very low; 5 = very high). |

The questionnaire included standard socio demographic variables and six questions to measure general attitudes toward technology (e.g., "Technology is a danger for humans and their environment," "Technology makes life more comfortable"). Participants answered the questions using a value between 1 ("don't agree at all") and 5 ("agree absolutely") [50].

## AUTHORS

**Gianfranco Elena** is a doctoral researcher at Glasgow University. He has been working as CTO and CIO in the Italian Army and NATO for more than 20 years. His expertise is in cybersecurity and risk assessment of cloud computing. His research contributes to improve the decision making process about the adoption of Cloud computing in Government organizations.

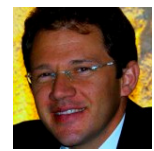

**Professor Chris Johnson** is Head of Computing Science at Glasgow University. His research increases the resilience of critical infrastructures. He is a software specialist on the SESAR scientific board advising the European Commission on the future of Air Traffic Management. He also focuses on the interactions between safety and security - for example, developing techniques so that we can safely close down a civil nuclear reactor even after malware has been detected.

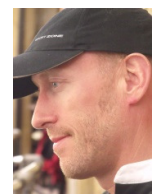